\documentclass[runningheads]{llncs}
\usepackage{graphicx}
\usepackage{enumitem}
\usepackage{amsfonts,lipsum}
\usepackage{tikz}
\usepackage{pgf-umlsd}
\usepackage{subcaption}
\usepackage[justification=centering]{caption}
\usepackage{amsmath}
\usepackage{tabularx}
\usepackage{ragged2e} 
\usepackage{hyperref}
\usepackage{academicons}
\usepackage{xcolor}
\usepackage{etoolbox}

\usepackage{siunitx}
\usepackage{xspace}
\usepackage{multirow}
\usepackage{afterpage}
\usepackage{svg}
\usepackage{footmisc}

\usepackage{svg}

\newcommand{\popularitylist}{\texttt{Popularity List}\xspace}
\newcommand{\mixnetworkvotes}{\texttt{Voting Mix Network}\xspace}
\newcommand{\vote}{\texttt{Vote}\xspace}
\newcommand{\votes}{\texttt{Votes}\xspace}

\title{DNS in the Time of Curiosity: A Tale of Collaborative User Privacy Protection}
\titlerunning{A Tale of Collaborative User Privacy Protection}

\author{Philip Sj{\"o}sv{\"a}rd \and Hongyu Jin \and Panos Papadimitratos}
\authorrunning{P. Sj{\"o}sv{\"a}rd \and H. Jin \and P. Papadimitratos}
\institute{KTH Royal Institute of Technology, Stockholm, Sweden\\ Networked Systems Security Group \\ \url{https://www.eecs.kth.se/nss} \\
\email{\{sjosv,hongyuj,papadim\}@kth.se}}

\begin{document}

\maketitle          

\begin{abstract}
The Domain Name System (DNS) is central to all Internet user activity, resolving accessed domain names into Internet Protocol (IP) addresses. As a result, curious DNS resolvers can learn everything about Internet users' interests. Public DNS resolvers are rising in popularity, offering low-latency resolution, high reliability, privacy-preserving policies, and support for encrypted DNS queries. However, client-resolver traffic encryption, increasingly deployed to protect users from eavesdroppers, does not protect users against curious resolvers. Similarly, privacy-preserving policies are based solely on written commitments and do not provide technical safeguards. Although DNS query relay schemes can separate duties to limit data accessible by each entity, they cannot prevent colluding entities from sharing user traffic logs. Thus, a key challenge remains: organizations operating public DNS resolvers, accounting for the majority of DNS resolutions, can potentially collect and analyze massive volumes of Internet user activity data. With DNS infrastructure that cannot be fully trusted, can we safeguard user privacy? We answer positively and advocate for a user-driven approach to reduce exposure to DNS services. We will discuss key ideas of the proposal, which aims to achieve a high level of privacy without sacrificing performance: maintaining low latency, network bandwidth, memory/storage overhead, and computational overhead.

\end{abstract}

\section{Introduction}

The Domain Name System (DNS) translates domain names into information necessary for the functionality of other protocols, most commonly Internet Protocol (IP) addresses \cite{dns_rfc},\cite{dns_resolver_cloudflare}. DNS resolvers respond to client DNS queries with a locally cached answer/translation or by querying an authoritative name server, located by recursively querying a distributed hierarchy of name servers \cite{dns_rfc},\cite{dns_resolver_cloudflare}. To prevent manipulation of name server responses, resolvers can authenticate them using Domain Name System Security Extensions (DNSSEC) \cite{dnssec}. Meanwhile, user-resolver queries can be encrypted using, for example, DNS over TLS (DoT) \cite{dot_rfc}, DNS over HTTPS (DoH) \cite{doh_rfc}, or DNSCrypt \cite{dnscrypt_protocol}. However, user privacy is not protected this way; DNS queries still reveal user web browsing activity and thus user interests to resolvers \cite{dns_privacy_recommendations_rfc}.

Granted, resolvers with a strong reputation for privacy and transparency can be chosen, especially if they support encryption of client DNS queries. Users could even try to rotate among multiple such resolvers to reduce their exposure to any one entity. Nevertheless, privacy policies (e.g., do not log specific user information), when in place, must be trusted, as users can hardly verify them alone. More importantly, public resolvers centralize DNS query handling: public resolvers account for nearly \SI{60}{\percent} of global name server DNS traffic \cite{dns_resolver_relevance}; one of them, i.e., Google DNS, account for approximately \SI{30}{\percent} \cite{dns_resolver_relevance}. Generally speaking, individual organizations handling potentially massive amounts of user data raise concerns about the potential misuse of sensitive user information. Thus, large-scale adoption of public resolvers, despite the services they offer, notably fast and reliable resolutions, is a \emph{double-edged} sword.

DNS queries can be anonymized by introducing a relay server that separates the sender identity from the query content \cite{odoh_rfc},\cite{anonymized_dnscrypt}. In such a setup, the resolver only sees the query content, and the relay only sees the sender IP address. However, this approach has limitations, as a colluding relay and resolver can easily re-associate the separated information, undermining user privacy.

To achieve stronger anonymity, anonymous networks such as Tor \cite{tor_paper},\cite{dns_over_tor} offer a solution, although they come with the drawback of significantly increased latency \cite{tor_performance},\cite{i2p_performance},\cite{anonymized_dnscrypt}. To improve privacy without negatively impacting the user experience, one approach is to use chaff queries: decoy traffic designed to obscure real queries \cite{dns_range_queries},\cite{dns_broadcast_range_queries_mix_zones}. However, this method is subject to high bandwidth usage and the risk of poor decoy quality \cite{dns_range_queries_pattern_analysis},\cite{dns_tracking}. A more promising alternative is to pre-download a list of popular, frequently resolved DNS records -- entirely avoiding exposure to external resolvers for the majority of queries, resolving them locally with minimal latency -- and resolving any remaining unpopular ones by querying an external resolver using a mix network \cite{dns_broadcast_range_queries_mix_zones}. To ensure that the answers, e.g., IP addresses, of the included records are up-to-date, users remain connected to the server that provided the list to receive incremental updates \cite{dns_broadcast_range_queries_mix_zones}. The server continuously re-queries the records as their Time-To-Lives (TTLs) expires, broadcasting any found changes (e.g., changes in IP addresses) to connected users. However, key questions remain: How do we instantiate such a list of popular records according to user interests in a privacy-preserving and secure manner? How do we ensure the relevance of the list as user interests change over time? How do we mitigate the excessive communication overhead from incrementally updating the list due to name servers employing aggressive DNS load-balancing schemes that continuously change record answers? How do we reduce the memory/storage overhead of such a list of popular records?

Our proposal is based on a two-fold approach for resolving DNS queries. First, users attempt to locally resolve queries using a pre-fetched \popularitylist containing the most popular DNS records. Naturally, the list should reflect the interests of its users, without undermining their privacy. Additionally, one cannot assume that all users are benign when aggregating their interests. Our proposal is to have clients/users contribute in a privacy-preserving manner to the \popularitylist creation using a specialized \mixnetworkvotes. The \popularitylist is maintained and distributed to clients/users by a not trusted, honest-but-curious public server. If a client cannot resolve a query using the \popularitylist, they can query an external DNS resolver using, for example, a public anonymous network (e.g., Tor).

Using cached data to improve privacy is not a novel concept. Instead, our contribution is the development of methods to anonymously and securely infer user interests to construct a relevant cache (i.e., the \popularitylist) and to efficiently distribute incremental updates to it. By designing an efficient voting mechanism and scheme to broadcast large volumes of incremental record updates to clients, DNS privacy can be achieved with low latency and network bandwidth overhead.

\section{System and Adversary Model}
\label{sec:system_and_adversary_model}

The system consists of three different entity types, ranging from \emph{honest-but-curious} -- entities interested in obtaining user information but not in interrupting the DNS protocol -- to curious and potentially malicious ones. These entities may collude with each other to achieve their goals. The entities are:

\begin{itemize}
    \item \textbf{A public server:} Central to our proposal, this server maintains and distributes the \popularitylist and its incremental updates. It is assumed to be honest-but-curious, meaning it correctly provides its advertised services to ensure the functionality of the DNS protocol, but remains interested in tracking user activities (in particular, client-resolved DNS records). The server may occasionally drop or alter packets, as irregular occurrences do not interrupt the general functionality of the DNS.
    \item \textbf{Existing DNS and Internet infrastructure:} These include public resolvers, name servers, Internet Service Provider (ISP) network infrastructure, etc. These entities are honest-but-curious, similar to the central public server.
    \item \textbf{Users/Clients:} They are considered curious and potentially malicious. They may attempt to disrupt the availability of the DNS protocol or infer user information by altering, dropping, or injecting packets.
\end{itemize}

\section{Problem Statement}
\label{sec:system_model_and_problem_statement}

Our goal is to minimize the exposure of Internet user activity to DNS services, which may collude and share traffic logs to link DNS queries to their user (e.g., IP address) origin. Unlike widely deployed privacy solutions that generally have a linear trade-off between privacy and performance, we aim to achieve privacy comparable to anonymous networks while maintaining inconsequential overhead, i.e., latency, bandwidth, computational, and memory/storage overhead. All while accounting for the ever-changing nature of DNS record data.

\section{Our Scheme}
\label{sec:scheme}

The \popularitylist central to our scheme contains the $N_{popular}$ most popular DNS records, as well as any intermediate CNAME records (records referencing/pointing to another domain) found among those $N_{popular}$. The list is maintained by a central public server (the public server in Section~\ref{sec:system_and_adversary_model}) and distributed to all clients/users within the system, which in turn can use it to resolve most DNS queries locally and anonymously. Given the finite size of the \popularitylist, DNS records not found in it are resolved using other privacy-preserving schemes, such as anonymous networks (e.g., Tor). Within the context of the \popularitylist, we call such schemes \emph{fallback DNS protocols}. The key feature of the \popularitylist is not necessarily to provide privacy per se, but instead to improve the user experience of otherwise costly privacy-enhancing schemes (e.g., Tor).

The \popularitylist memory and storage overhead is generally small due to the long-tail distribution of DNS record popularity, with a small number of records accounting for the majority of queries \cite{dns_broadcast_range_queries_mix_zones}. We mirror the hierarchical tree structure of domain names to further reduce the \popularitylist size, compared to storing domain-answer pairs line by line. For example, accumulating all \textit{.com} domains under a single section/header allows \textit{example.com} to be represented by \textit{example}, effectively compressing domain names.

Given that DNS records can change at any point (e.g., changes in IP addresses), the central public server ensures that the \popularitylist is up to date. That is, any DNS record whose TTL expires is immediately re-queried by the server to receive an up-to-date response from a DNS resolver or name server. If the new response differs from the previous one, the server will broadcast the incremental update to all connected clients. To reduce network overhead, we accumulate several updates and send them in the same message, enforcing only a small minimum TTL (e.g., \SI{60}{\second}) inconsequential to the DNS protocol.

\subsection{Efficient Broadcast of Incremental Updates}
\label{sec:incremental_updates}

DNS load-balancing schemes are widely used to distribute users across servers by continuously changing the IP addresses of DNS records. We find that these records often have TTLs as short as \SI{30}{\second}, resulting in high network usage to incrementally update the \popularitylist, as is the case in \cite{dns_broadcast_range_queries_mix_zones}. To reduce the amount of information transmitted in each incremental update, our \popularitylist maintains a pool of answers (typically IP addresses) that at some point have been relevant for at least one record in the list. Instead of transmitting the new answer in its entirety, incremental updates can instead reference/point to the relevant answer in the already cached pool of potential answers by referencing its order of appearance relative to the current one. Similarly, the record whose answer should be updated can be identified by its order of appearance in the \popularitylist, significantly reducing the incurred network overhead of incremental updates.

\subsection{Voting on the Contents of the Popularity List}

The set of DNS records included in the \popularitylist is continuously updated according to anonymous \votes submitted by users via a specialized \mixnetworkvotes. These \votes, which reflect user queries in the last $t_{refresh}$ seconds, are submitted simultaneously by all clients at set $t_{refresh}$ intervals, allowing our server to periodically refresh the list according to the interests of its users. Any resulting \popularitylist updates are broadcasted by the server to all connected clients.

Unlike traditional mix networks, the \mixnetworkvotes does not utilize third-party servers for mixing its packets (i.e., \votes). Instead, connected clients act as mix nodes, collectively anonymizing the \votes. Furthermore, our server also relays mix network packets between each mix node (client) for a more efficient and implementable approach compared to direct node-to-node communication, avoiding firewalls and Network Address Translations (NATs). But it also does so to monitor incoming and outgoing \votes at each mix node, to ensure that no entity submits more \votes than allowed or injects additional ones. In a traditional mix network, a central node monitoring all connections would undermine its ability to anonymize packets due to possible timing attacks. However, the \mixnetworkvotes is built around all clients casting their \votes at the same time and mixing them in distinct \emph{shuffling} rounds, alleviating any such concerns. The overarching process steps are:

\begin{enumerate}
    \item Each client iteratively encrypts each of their \votes using different sets of public keys corresponding to a random set of mix nodes (other clients) in the \mixnetworkvotes. Every \vote now has a designated (unique) path through the \mixnetworkvotes.
    \item The server receives all initial \votes from all clients, sorts them into batches based on their next-hop destination, and redistributes them.
    \item Upon receiving its batch of \votes, each client decrypts the top layer of encryptions using their private key and randomly shuffles the order of the \votes.
    \item The server then collects all shuffled batches from all clients and redistributes them, repeating the process.
    \item The process continues until all \votes are fully decrypted and no longer are forwarded, allowing the \popularitylist to be updated with the now anonymized \votes.
\end{enumerate}

To prevent Sybil attacks, where attackers use a large number of identities/devices to gain increased influence over the \popularitylist and/or undermine the anonymity of the \mixnetworkvotes, users must register in advance, e.g., with a phone number, and authenticate using acquired credentials upon connecting to the server. It is important to note that we, much like anonymous networks (e.g., Tor), anonymize user activity, not their identity. For example, a user's ISP can observe that they are connected to a Tor entry node.

Furthermore, to ensure that the server does not perform similar attacks by impersonating a large set of clients, users can validate the identity of peer clients using certificates obtained from an external Public Key Infrastructure (PKI) during registration. Specifically, these certificates are used to verify the public keys of clients participating in the \mixnetworkvotes. To hide user identities, one may use ephemeral pseudonymous client certificates \cite{vpki_1},\cite{vpki_2}.

\section{Preliminary Results}

To simulate a real-world application of our scheme, we use a dataset of DNS traffic collected from a campus network over ten days \cite{dataset}, which peaks at around \num{4000} active hosts \cite{dataset}. The \popularitylist is instantiated by the queries of the first day of the dataset, and then updated once an hour through a communal voting round using the \mixnetworkvotes ($t_{refresh} = \SI{3600}{\second}$). For each hourly voting round, clients cast \votes for up to ten unique DNS records, where each \vote is based on the DNS queries the client/user resolved during the last $t_{refresh}$. Each user DNS query has a \SI{30}{\percent} probability of generating a \vote. The voted-on records are ranked based on their weighted number of occurrences, $w_m$,
\begin{equation}
    w_m = 0.1 n_m + 0.9 w_{m-1} \,,
\end{equation}
during voting round $m$, based on the number of received \votes, $n_m$.

Table~\ref{tab:hit_ratio} shows the mean hourly hit ratio of the \popularitylist during the last nine days of the dataset. The high hit ratio implies that the vast majority of user DNS queries are resolved locally by the \popularitylist, effectively substantially reducing: the mean latency of the fallback DNS protocol compared to using it as the sole DNS protocol, and the number of queries exposed to the DNS resolver. These results are comparable to the \SI{83.9}{\percent} and \SI{94.5}{\percent} seen in \cite{dns_broadcast_range_queries_mix_zones} for a list of size \num{10000} and \num{100000}, respectively. \footnote{The compared \textit{Optimal TopList} from \cite{dns_broadcast_range_queries_mix_zones} is not implementable in real-world scenarios, as it is constructed from future user queries; users whose queries it cannot reliably sample from in general, as there is no voting (or equivalent) mechanism.}

\begin{table*}
    \begin{center}
        \caption{Mean hourly hit ratio of our \popularitylist.}
        \label{tab:hit_ratio}
        \begin{tabular}{|l|l|}
            \hline
            \textbf{$N_{popular}$} & \textbf{Hit ratio} \\ \hline
            \num{10000} & \SI{91.4}{\percent} \\ \hline
            \num{25000} & \SI{94.4}{\percent} \\ \hline
            \num{100000} & \SI{96.2}{\percent} \\ \hline
        \end{tabular}
    \end{center}
\end{table*}

\begin{figure}[t!]
    \centering
    \includegraphics[width=0.7\linewidth]{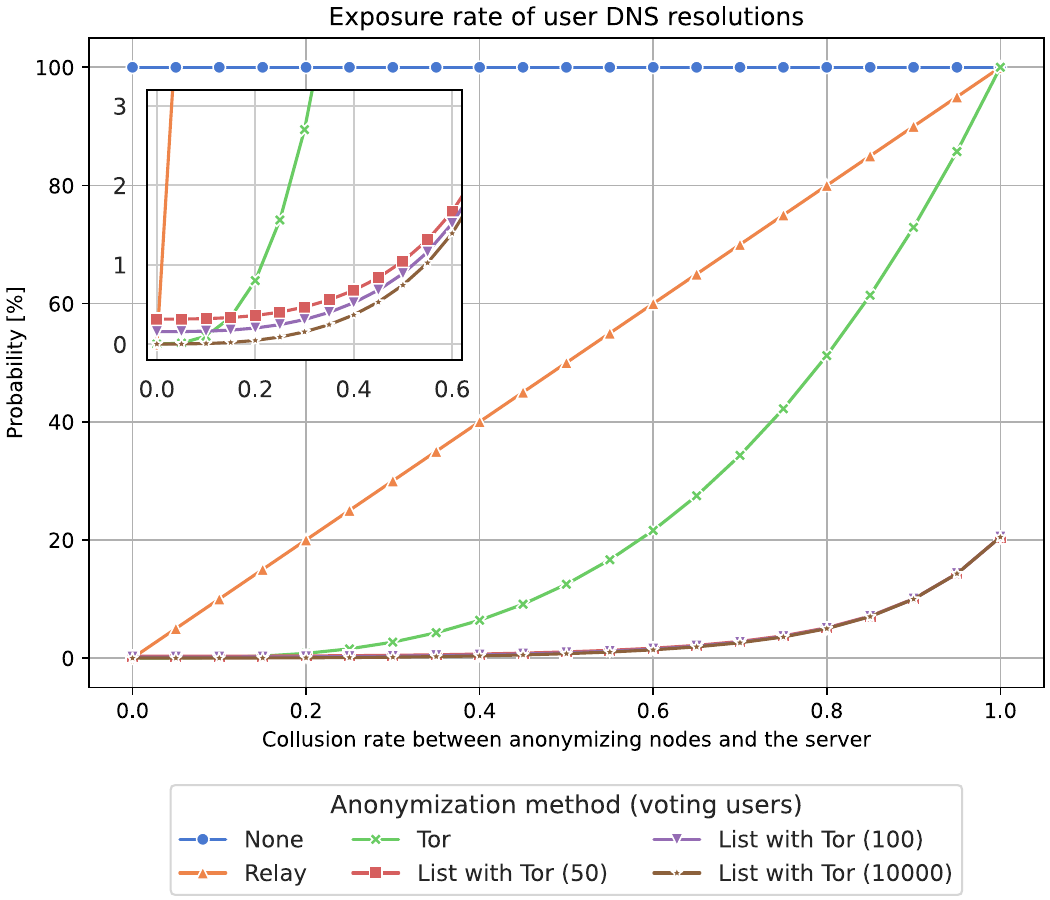}
    \caption{Exposure rate -- the probability that a DNS resolver or our public server correctly guesses the origin of a user-performed DNS query -- for various collusion rates between the DNS resolver/public server and other nodes. Tor uses three nodes.}
    \label{fig:exposure_rate}
\end{figure}

For our scheme, we use the standard Tor configuration (three nodes) as the fallback protocol for resolving queries not in the \popularitylist. Figure~\ref{fig:exposure_rate} shows the \emph{exposure rate} of our scheme (using $N_{popular} = \num{25000}$ and ten \mixnetworkvotes shuffling rounds per voting round) with Tor as fallback, comparing it with widely deployed DNS anonymization methods. The exposure rate is calculated as the probability that a DNS resolver or our public server correctly guesses/infers the origin of a user-performed DNS query, including the potential exposure of converting them to \votes; unlike the likely large anonymity set of users for widely deployed schemes, the pool of voters could potentially be small enough to have privacy implications. Compared are various collusion rates between the resolver/public server and anonymizing relays (e.g., single DNS query relay, Tor nodes, other clients, etc.). Overall, our scheme achieves a low exposure rate, even at high collusion rates. The scheme manages to improve the exposure rate, and, importantly, also the mean latency of Tor, as \SI{94.4}{\percent} of queries are locally cached by the \popularitylist. The exception is a slight increase in the exposure rate for low collusion rates due to the potentially limited pool of voting users, allowing \votes to be associated with the current pool of users. Interestingly, the performance of \num{50} and \num{10000} voters is comparable, indicating a high level of privacy despite potential cases of poor participation in the voting system.

\begin{figure}
    \centering
    \includegraphics[width=0.7\linewidth]{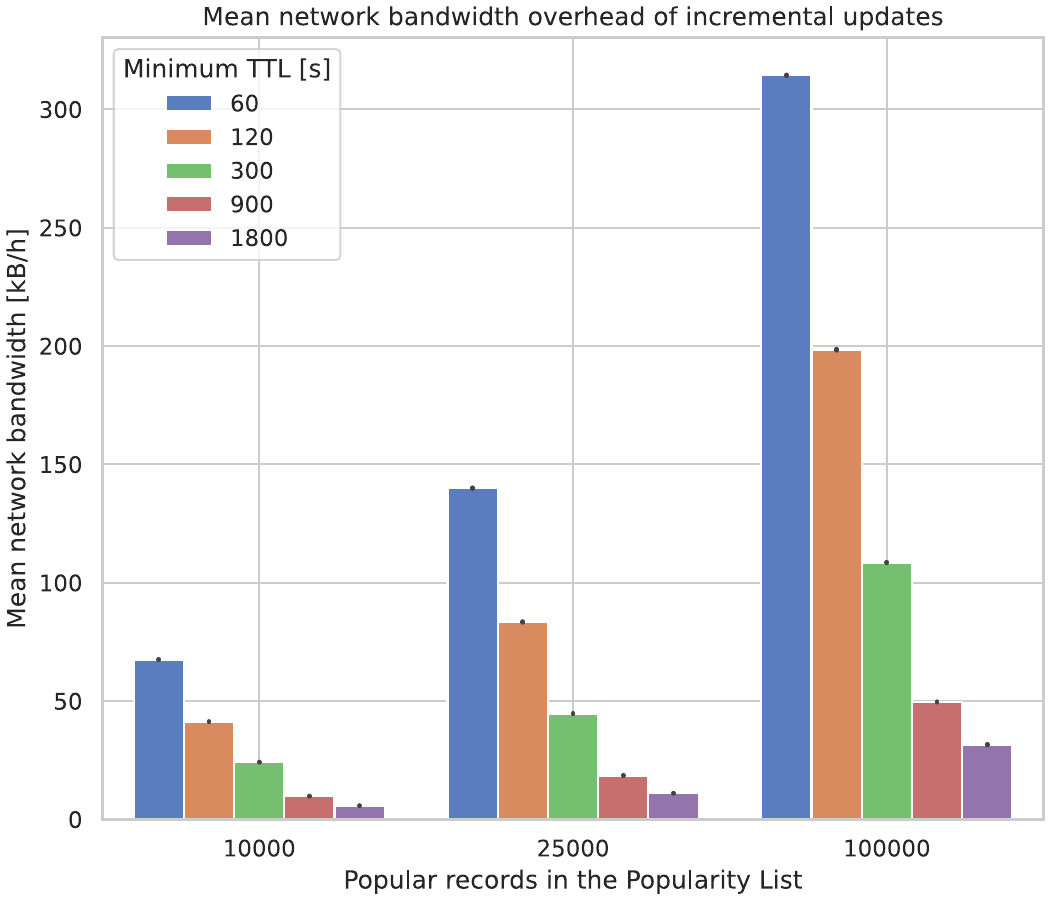}
    \caption{Mean client network bandwidth overhead for incrementally updating the \popularitylist, applying compression (Python's \emph{zlib} library) whenever applicable, for various $N_{popular}$ and minimum record TTLs.}
    \label{fig:incremental_updates}
\end{figure}

Figure~\ref{fig:incremental_updates} shows the mean client network bandwidth overhead for incrementally updating the \popularitylist when, for example, the IP addresses of its DNS records change, for various enforced minimum record TTLs (reflecting the time between transmitted incremental updates). As shown, the overhead of incremental updates is minimal. For comparison, \cite{dns_broadcast_range_queries_mix_zones} achieves around \SI{1500}{\kilo\byte/\hour} (\SI{3.3}{\kilo\bit/\second}) for a list of size \num{10000}: \num{20} times that of the \SI{67.5}{\kilo\byte/\hour} (\SI{0.15}{\kilo\bit/\second}) our scheme achieves with the \SI{60}{\second} minimum TTL configuration. These improvements are achieved despite, according to preliminary testing, DNS records likely changing more frequently today compared to 2011 due to noticeable reductions in average DNS record TTLs.

\section{Conclusion}

We consider preparedness to safeguard user privacy based on user collaboration, to avoid relying on the assumption that DNS resolvers and other Internet infrastructure are not curious. The objective is to ensure low-latency, low-bandwidth, and reliable DNS query resolution with strong user privacy protection. We propose a user-driven \mixnetworkvotes, allowing users to anonymously vote on the contents of a securely instantiated, pre-downloaded \popularitylist of frequently resolved DNS records; a list with an answer cache mechanism to enable efficient incremental updates. In this way, user exposure to public resolvers is minimized while remaining efficient and compatible with the current DNS ecosystem.

\section{Acknowledgments}

This work was supported in parts by the Strategic Research Foundation (SSF).

\bibliographystyle{splncs04}
\bibliography{sample-bibliography}

\end{document}